\documentstyle[prd,aps,preprint,epsfig]{revtex}
\input epsf
\begin{document}
\title{ The Critical Points of Strongly Coupled 
Lattice QCD at Nonzero Chemical Potential. }
\vskip .6 truecm
\author { Ian~M.~Barbour\thanks{UKQCD Collaboration} 
and Susan~E.~Morrison$^{*}$} 
\address{ Department of Physics and Astronomy,University of Glasgow,
Glasgow G12 8QQ, Scotland} 
\author {Elyakum G.~Klepfish$^{*}$} 
\address {Department of Physics, King's College London, London WC2R 2LS, UK} 
\author {John~B.~Kogut } 
\address { University of Illinois at Urbana--Champaign,
1110 West Green Street, Urbana, Illinois 61801-3080} 
\author {Maria--Paola~Lombardo}
\address {Zentrum f\"ur interdisziplin\"are Forschung,
  Universit\"at Bielefeld,  D-33615 Bielefeld, Germany}
\maketitle

\begin{abstract}
We study QCD at non--zero quark density, zero temperature,
infinite coupling  using the Glasgow algorithm.
An improved  complex zero analysis gives a critical point $\mu_c$
in agreement with  that of chiral
symmetry restoration computed with strong coupling expansions,
and monomer--dimer simulations. 
We observe, however, two unphysical critical points: the onset for
the number density $\mu_o$, and $\mu_s$ the saturation threshold,  coincident
with pathological onsets observed in past quenched QCD
calculations. An analysis of the probability  distributions 
for particle number supports our physical interpretation 
of the critical point $\mu_c$,
and  offers a new intepretation of $\mu_o$,  
which confirms its unphysical nature.
The perspectives for future lattice QCD calculations of the properties of dense
baryonic matter are briefly discussed. 
\end{abstract}

\newpage
\setcounter{page}{1}
\pagenumbering{arabic}

\section{Introduction}

Numerical simulations of lattice QCD at high temperatures 
are making quantitative 
theoretical predictions  which 
will be confronted with experiments at RHIC and LHC \cite{latrev}.
However, numerical simulations of QCD in an environment rich in baryons lags far behind.
Phenomenologically we know that nuclear matter can exist 
up to a density of four times ordinary matter in
neutron stars, and that higher density will eventually 
induce deconfinement and chiral
symmetry restoration because of asymptoptic freedom. 
Current estimates from phenomenological
nuclear models \cite{KKL97} place the critical chemical potential
between 1000 and 1600 Mev, and the critical baryon density between
twice or twenty times that of ordinary nuclear matter.

As it is well known, the reason behind this poor 
knowledge
is the lack of a reliable calculational scheme
for lattice QCD at high baryon densities \cite {BMKKL97}. 
A solid theoretical formulation for finite density QCD was made
ten years ago \cite{BMSW83}, \cite{HK83}, but,
since the resulting action is complex, probabilistic simulation methods 
fail. Early approaches considered  
the quenched approximation,  which omits
the complex fermion determinant, but it produced unphysical 
results \cite{BBD86}--\cite{LKS96}.
To obtain reliable results the determinant should be included,
and the Glasgow method has been proposed to tackle this challenge
\cite{BDS88}--\cite{SEM97}.
Although we are ultimately interested in the weak coupling, continuum
limit, the strong coupling limit of QCD is attractive for
several reasons : 1. There are
analytic results coming from the strong coupling 
expansion \cite{DHK85}, \cite{IK85}, \cite{BDP92} and numerical
results from monomer--dimer simulations \cite{KM89}, 2. The theory confines and
spontaneously breaks chiral symmetry. In this paper we will use these features
of strongly coupled lattice QCD to test and shed light on simulation
methods which could be used at any coupling \cite{MPL97}.

This paper is organized into two main Sections, Method and Results.

{\it Method} is part review, and part illustration of our method of
analysis. We first review  the Glasgow method, and the relevant 
observables (II.A). We continue by illustrating 
some features which will help our numerical analysis. 
We will first discuss (II.B) the pathologies found in calculations on
isolated configurations. 
In Section II.C we will  discuss how the Glasgow
algorithm can escape from these single configuration
pathologies, and build the 
physical signatures for the critical point $\mu_c$ where
chiral symmetry is restored.

In {\it Results}, after discussing some generalities
of the generation of the gauge ensemble (III.A), we show
that the Glasgow method results for the Baryon current inherit 
some of the quenched/single configuration pathologies (III.B).
Nevertheless,  we will succesfully measure the critical
chemical potential $\mu_c$ (III.C) and we will discuss in detail
the interplay of successes and failures of the results.
Comprehensive results for an extended range of masses
are given in Section III.D.
In Section III.E we will present
an alternative reanalysis of the critical region which will
use probability distributions. 
We will confirm there the estimate of the critical point
$\mu_c$, and we will offer a new intepretation
of the pathological region.

We conclude with a brief summary and discussion.

\section{Method}

Given the failure of the quenched approximation to deal with the problem
of the chiral phase transition at high quark density, 
the natural conclusion is that dynamical quark simulations are essential.
However, the complex measure of the functional integration with
nonzero chemical potential poses a severe problem for such simulations.
One simulation method which circumvents this problem
is based on the expansion of the
Grand-Canonical Partition Function (GCPF) in powers of the fugacity. 
The GCPF, $(Z)$, can be written as the ensemble average 
$\langle \frac {|M(\mu,m)|}{|M(0,m)|} \rangle$ where $|M(\mu,m)|$ is 
the fermion
determinant at chemical potential $\mu$ and quark mass $m$ in lattice units
(lattice spacing $a=1$), i.e.
\begin{equation}
Z={{\int [dU][dU^\dagger] |M(\mu,m)|e^{-S_g[U,U^\dagger]}}\over
{\int [dU][dU^\dagger]|M(0,m)|e^{-S_g[U,U^\dagger]}}}
\end{equation}
where $U$ are matrices representing the gauge degrees of freedom and
$S_g$ is the standard Wilson gauge action. The fermion matrix $M$ is that
describing 4-flavours of staggered fermions.

The fermionic determinant can be expressed explicitly 
as a function of $\mu$ by
\begin{equation}
\det(M(\mu,m)) = e^{-3\mu n_s^3 n_t} \det(P-e^{\mu})
\label{for:detdet}
\end{equation}
The lattice size is $n_s^3 n_t$
and $P$ is the propagator matrix (independent of $\mu$)\cite{Gi86}
\begin{equation}
P=\left(\begin{array}{cc}
-GV & V \\
-V  & 0
\end{array} \right)
\end{equation}
where $G$ contains all the spacelike gauge links and the quark
bare mass, and $V$ all the forward timelike links 
of the fermionic matrix M.

$\det (M(\mu, m)$ can be  computed in  a basis 
where the propagator matrix is diagonal
\begin{equation}
\det(M(\mu,m)) = 
e^{-3\mu n_s^3 n_t}\prod_{k = 1}^{6 n_s^3 n_t}(\lambda_k-e^{\mu})
\label{for:deteig}
\end{equation}
We recognize that the zeros of the determinant in the
$e^{\mu}$ plane are the eigenvalues
of the propagator matrix.
The symmetry of the eigenvalues of 
the propagator matrix  
$\lambda_{k+j} = e^{i2\pi j/n_t} \lambda_k$ for 
$j =0$ to $n_t - 1$, together with the polynomial decomposition
\begin{equation}
\prod_{j=0}^{n_t - 1} (e^{i 2 \pi j / n_t} \beta - x) = 
 (\beta^{n_t} - x^{n_t})
\label{for:poly}
\end{equation}
yields the equivalent representation 
\begin{equation}
\det(M(\mu,m)) = e^{-3 \mu n_s^3 n_t} 
\prod_{k=1}^{6 n_s^3} (\lambda_k^{n_t} - e^{\mu n_t})
\label{for:deteigT} 
\end{equation}
and dictates the general structure of the characteristic polynomial
$\det(M(\mu, m))$ 
\begin{equation}
\det(M(\mu, m)) = \sum_{k =-3 n_s^3}^{3 n_s^3} b_i e^{k \mu n_t } 
\end{equation}
Note the dependence on $\mu$ is now via the fugacity $f = e^{\mu n_t}$.

Hence, measurement
of the average of the characteristic 
polynomials  (normalised by $|M(0,m)|$) 
in the ensemble generated at 
update mass $m$ and $\mu=0$ will give
$Z(\mu,m)$ explicitly as a function of $\mu$ at that mass.

This representation leads to a polynomial expansion of $Z(\mu)$ 
in powers of the fugacity whose coefficients are functions 
of the gluonic fields.
\begin{equation}
Z(\mu) =  \sum_{k =-3 n_s^3}^{3 n_s^3} <b_i> e^{k \mu T } 
= \sum_{k =-3 n_s^3}^{3 n_s^3} Q_k f^k
\label{for:basic} 
\end{equation}

This expansion is just that of the GCPF expanded in terms
of the canonical partition functions, (CPF's), for a fixed number of 
quarks (anti-quarks)on the lattice.
Thermodynamical averages,
which can be calculated as logarithmic derivatives of the GCPF, are then 
given explicitly as functions of $\mu$.

The relative
value of the CPFs can characterise the
properties of the system as well. For example, the relative weight
of the triality--bearing to the triality--zero CPFs
can signal whether the system is in the confined or deconfined phase. In 
the confined phase the ensemble average of the triality--bearing CPFs must 
be zero. This leads to
\begin{equation}
Z(\mu) =  \sum_{k = n_s^3}^{n_s^3} Q_{3k} f^{3k}
\label{for:basic3} 
\end{equation}

One can also explore the phase structure 
of the simulated system by examining the distribution of the
zeros of the GCPF in the complex chemical potential (or fugacity) 
plane \cite{YANGLEE} \cite{IPZ83}.
These zeros correspond to the singularities of the thermodynamic potential and
will converge in the thermodynamic limit, 
($L \rightarrow \infty$)
towards any critical $\mu$ in the physical domain. 

In the following we also show that one can regard the zeros of the averaged
characteristic polynomial as the ``proper'' ensemble average of the 
eigenvalues of the propagator matrix.
This interpretation of the zeros as reflecting the ensemble average of the 
eigenvalues could be important in the interpretation of ``unexpected'' onset
chemical potentials in ensembles of limited statistics.

\subsection{Observables}

The starting point of our  analysis is the
GCPF $Z$ computed with the Glasgow algorithm. 
Our raw data are the CPF's $Q_N$
and our basic observables are the particle number density
and the zeros of the GCPF in the complex fugacity plane.

Most of our discussions will consider the current 
$<J_0(\mu, m) >$ or equivalently the particle number density, defined as
\begin{equation}
<J_0(\mu, m)> = \frac{1}{V} \frac{\partial \ln(Z(\mu, m))}{\partial \mu} 
= \frac{1}{V} \frac{\partial \ln <\det(M(\mu, m))>}{ \partial \mu} 
\label{for:j0}
\end{equation}

Singular behaviour of the current can result from singularities in the density
of baryonic states (particularly apparent in the zero-temperature
limit). These singularities could be purely lattice
artefacts and vanish in the continuum limit. However, they may instead reflect
continuum spectral features, such as gaps in the spectrum or abrupt
changes in the dispersion relation of the baryonic excitations. A chiral
phase transition is one such possibility. A
spectrum of chirally symmetric baryonic excitations will follow a gapless
relativistic dispersion relation, contrary to the dispersion of particles with
broken chiral symmetry. If the disappearance of the
mass gap occurs together with the deconfinement transition, quark
states will emerge instead of collective colourless baryonic excitations.
Thus, the $\mu-$dependence of $J_0$ should determine
the phase structure of dense baryonic matter, an alternative
to the evaluation of the chiral condensate.

Differentiating the action with respect to $\mu$ reveals
the operator form of the charge, and one sees that
the current is the expectation value of the number of paths
through the links in the time direction \cite{BMSW83}. 
In this sense the current
can be defined on isolated configurations, where it
reduces to
\begin{equation}
J_0^i (\mu, m) = \frac{1}{V} \frac{\partial \ln(\det(M(\mu,m)))}{ \partial \mu}
\label{for:j0i}
\end{equation}
In the quenched ensemble $\ln(\det(M))$ is differentiated
{\it before} taking the statistical average:
\begin{equation}
<J_0>^q (\mu, m)= \frac{1}{V} 
\left<\frac{\partial \ln(\det(M (\mu,m)))}{ \partial \mu} \right > 
\label{for:j0que}
\end{equation}
and we recognize that
\begin{equation}
<J_0(\mu,m)>^q = <J_0^i(\mu,m)>
\label{for:j0quei}
\end{equation}
In the following the $\mu$ and $m$ dependence will be left implicit
wherever this does not create ambiguities.

\subsection{Failures on isolated configurations, and the quenched model}

The early work by Gibbs \cite{Gi86} made it clear that the behaviour
of some observables measured
on isolated configurations at finite density
can be pathological. Since the 
analysis of isolated configurations is a necessary
step in any lattice simulation, the impact
of his result may be broader that its original
motivation-- to understand the
pathologies of the quenched approximation. 

Our renewed interest was prompted by two considerations.
First, our results presented in Section III.A below
show clear relics of the quenched pathologies discussed
in Gibbs's paper: the onset $\mu_o$ where the current $J_0$ departs 
from zero is
at half the pion mass. Second, published results on 
four fermion models \cite{HKK95} 
\cite{BMK96} do not have such pathologies.
However, both models share the same pattern of chiral symmetry
breaking, and both models have Goldstone modes. 
Why, then, is there a difference at finite density?
We decided to re--examine the behaviour of observables
on isolated configurations in order to test the Gibbs scenario
in a more general framework, and to gain some 
understanding of the process of statistical averaging
in the two models. This paper is devoted to QCD, and the results
for four fermion models will be presented elsewhere.

First consider the behaviour of the current
on isolated configurations.
$J_0^i$ follows from eqs. \ref{for:j0i}, \ref{for:deteig} 
\begin{equation}
J_0^i = -1 + \frac{1}{V} \sum_{i=1}^{6 V} z/( z - \lambda_i)
\label{for:j0_q0}
\end{equation}
(here and in the following we use Gibbs' notation $z = e^\mu$ \cite{Gi86}).
In the zero temperature case the sum over complex poles can be
conveniently done by contour integration, yielding:
\begin{equation}
J_0^i = \frac{1}{V} \sum_{1 < |\lambda_i| < e^\mu} 1.
\label{for:j0_q1}
\end{equation}

The threshold for the current $J_0$ on isolated 
configurations is triggered by the lowest zero 
of the determinant. In turn, 
the zeros of the determinant are given by the eigenvalues
of the Propagator Matrix that, as emphasized by Gibbs,
are controlled by the mass spectrum
of the theory. The argument,
which we briefly summarize for the sake of completeness, 
requires the calculation of the hadronic spectrum on replicated
lattices, i.e. lattices which have been strung together $d$ times
in the time direction and the limit $d \to \infty$ is taken,
in order to replace finite sums with contour integrations.
This procedure is justifiable at zero temperature. 

The expression for
the inverse of the fermion matrix, $G(t_1,t_2)$, on the replicated lattices
reads (slightly simplifying Gibbs' notation),
\begin{equation}
G(t_1,t_2) = \sum_k A_a \lambda_k ^{t_1 - t_2}
\label{for:prop}
\end{equation}
where the $A_a$ are the amplitudes which can be related to the
eigenvectors of the propagator matrix, and the $\lambda_k$ are
the corresponding eigenvalues.

Eq. \ref{for:prop} shows that the exponential 
decay of $G(t_1,t_2)$ at large time--like separation is
controlled by the eigenvalues of the propagator matrix. In other words, 
the eigenvalue spectrum calculated on isolated configurations
should be closely related to the physical mass spectrum.
In particular, Gibbs concluded that
the smallest mass state $m_\pi$ is related
to the lowest eigenvalue :
\begin{equation}
m_\pi = 2 \ln |\lambda_{\min}|
\end{equation}
This identification was clear in simulations done by Gibbs because
the pion propagator was very similar configuration by configuration 
although, strictly
speaking, masses are properties only of the statistical
ensemble.

Gibbs argument has been reformulated and verified by 
Davies and Klepfish \cite{DK91}.
Pathologies of isolated configurations,
the role of confinement and other issues, 
are also discussed in \cite{KLS95}, \cite{LKS96}. 
All of these works confirm that on isolated
configurations there is a singularity at a value of
the chemical potential close to half the pion mass.

Therefore, the results on 
isolated configurations are qualitatively different 
from those expected of the statistical ensemble.

Some of the problems with the quenched model 
can be understood from eq. \ref{for:j0quei} : the quenched current
is a simple average of the one--configuration--current,
and the quenched ensemble retains the pathological features
observed on isolated configurations.

\subsection{The statistical ensemble, and the full model}

We can now focus on the interplay between the 
one configuration/quenched 
results and ensemble results. How
can statistical averaging remove the 
problems observed on isolated configurations? Equivalently,
how can the Glasgow averaging discussed above
improve upon the quenched approximation?

Consider the fugacity expansion for $Z$, eq. \ref{for:basic}. 
By reinstating a factor $e^{3 n_s^3 n_t \mu}$  
we see that $Z$ is a polynomial
of degree $6n_s^3 n_t = 6V$ in the variable $z =e^\mu$.
$Z$ can then be written
in terms of its zeros $\alpha_i$ in the $z$ plane
\begin{equation}
Z = e^{3 V \mu} \prod_{i=1}^{6 V}( z - \alpha_i).
\label{for:zeta}
\end{equation}

Recall that 
$Z = <det (M)>$ and compare 
formulae \ref{for:zeta}  and \ref{for:deteig}  which we
rewrite here
\begin{equation}
\det M = e^{3 V \mu} \prod_{i=1}^{6 V}( z - \lambda_i) 
\nonumber
\end{equation}
We see that the zeros of the partition function 
are the ``proper'' ensemble average
of the eigenvalues of the fermionic propagator matrix, or,
equivalently, of the zeros of the determinant. 

Manipulations analogous to those of 
eqs. \ref{for:j0_q0}, \ref{for:j0_q1} lead to the current
\begin{equation}
J_0 = \frac{1}{V} \sum_{1 < |\alpha_i| < e^\mu} 1.
\label{for:jjj}
\end{equation}

Let's search for other critical points past the first onset.
From eq. \ref{for:jjj} we see that
discontinuities in $J_0$ are associated
with a high density of zeros on circles
with radius $e^{\mu_c}$ in the fugacity plane. 
More generally, the density 
of the modulos of zeros in the $e^\mu$ plane
is the derivative of $J_0$ with respect to $\mu$, 
i.e. the quark number susceptibility.
Interestingly, the relevant quantities controlling
the critical behaviour of the current are indeed 
the modulos of the complex zeros.

It is worth noticing that once the $Z3$ symmetry 
is enforced (eq. \ref{for:basic3})
\begin{equation}
Z = e^{3 n_s^3 \mu} \prod_{i=1}^{2 V}( z^3 - \beta_i).
\end{equation}
The zeros in the complex plane $z$ should then
come in triplets,
corresponding to cubic roots of certain complex 
numbers $\beta_i$. {\it In principle} (in practice
things can be very different!) the effect of 
the $Z3$ symmetry can simply amount to a
redistribution of  phases with no effect on the moduli.
That would not affect the critical 
behaviour, since the critical behaviour is triggered by the
moduli themselves. The unphysical
quenched onsets could certainly survive the $Z3$ symmetry 
of the full ensemble. 

The zeros of the partition function 
drive the critical behaviour of the full model  
as the zeros of the determinant drive the critical behaviour
of isolated configurations, hence of the quenched model. 
In the process of  going from
the zeros of the determinant to the 
zeros of the grand canonical partition function, the 
pathological results observed on isolated  configurations
should turn into  the physics of the  full model :
the fake critical points should disappear, 
the real phase transitions of the full model should emerge.

\section{ Results}

We present here our numerical results. All the background
material, when not explicitly referenced, 
can be found in the previous Section.

The generation of
the configurations of gauge fields 
is described in {\it The ensemble }

In {\it The number density}  we  review past results from the
quenched approximation, from the monomer--dimer simulation of
the full four flavor model and
from the strong coupling expansion of the four flavor model. 
We  present the results obtained with the Glasgow
method on various lattice sizes, and we highlight
the main similiarities and differences among the Glasgow
results and the above--mentioned ones. 
A discussion of finite size effects is presented as well.
In this and in the subsequent subsection,
the emphasis is on the presentation
of the main features of the results. We then limit ourselves 
to a discussion of one representative mass value, $m_q = .1$.

In {\it The determination of the critical point}  we focus on
the analysis of the complex zeros in the $e^\mu$ plane. 
We  contrast the pattern of zeros with that 
of the eigenvalues of the fermion propagator. 
We show how simulations of the full model
produce a clear signature for
the critical point $\mu_c$.

In  {\it Light and heavy masses} we
will present our complete set of results. The light masses
will offer information about the chiral limit. We 
confirm that our estimate $\mu_c$  remains constant, and
different from zero when $m \to 0$. We  demonstrate
the dependence of $\mu_c$ on the bare mass in the heavy quark
regime. In the summary plot 
the results are compared with the 
strong coupling expansion, and the monomer--dimer calculations.

{\it The analysis of the probability distribution} 
consists of a self-consistent
analysis of the critical region which  exploits the form of the 
probability distribution as a function of chemical potential. 
This analysis  further validates our estimate of the critical point
$\mu_c$ and, in addition,  offers a new intepretation
of the onset region $\mu_o$.

\subsection{The ensemble}

The Glasgow algorithm takes as input an ensemble
of configurations at zero chemical potential.
At infinite gauge coupling we can generate configurations
either with the usual hybrid MonteCarlo procedure or just
choosing random SU(3) matrices -- this corresponds to 
different normalization for the partition function.
First, we reproduced the results of Barbour,
Davies and Sabeur\cite {BDS88} , which show
that the reweighting actually works on a $2^4$ lattice
provided the statistics are high enough. 
Some preliminary runs were performed on a $4^4$ lattice
to check that the results were indeed independent of the algorithm
chosen for the generation of the configuration. We finally
selected a random generator which produces decorrelated 
configurations. 

We will present results
on a $6^4$ lattice for bare mass values ranging 
from .05 to 1.5 and on a $8^4$ lattice for masses .08 and .1. 
The number of gauge field configurations analyzed ranges from
a small sample of 25 on the $8^4$ lattice $m_q = .08$, 
$\simeq$ 100 configurations  on the same lattice, $m_q = .1$,
and several hundred configurations on the $6^4$
lattices.

\subsection{The number density}

We studied $m_q = .1$ on $6^4$ lattices where we can contrast the results
with those obtained 1. in the quenched case, 2. with
the monomer--dimer simulation, and 3. with the
analytic results of the strong coupling expansions.
Let's briefly review methods 2. and 3.

The monomer--dimer approach (valid only at infinite coupling) 
writes the strong coupling action
in a fashion suitable for computer simulations. It begins with the
standard lattice QCD action with four flavors of staggered fermions
and integrates out the completely disordered gauge fields. Confinement is
enforced exactly and the short-ranged interactions between fermions allow
the Grassman integrals to also be done exactly. The resulting action
can be interpreted graphically in terms of 'monomers' and 'dimers',
familiar constructions in statistical physics. This representation of the
theory is well suited for computer simulations since the dreaded
sign problem of the fermion determinant is not numerically 
significant in this basis (on small lattices).
Computer simulations at $m_q = .1$ 
\cite{KM89} show a sharp transition at $\mu = .69(1)$ where
the chiral condensate falls from its zero-$\mu$ value (essentially) to 
zero, and
the number density $J_0$ jumps from zero (essentially) to a fully occupied
lattice, $J_0 = 1.0$. These results agree with those of the traditional 
strong coupling expansion, as they should. 

The strong coupling expansion \cite{DHK85}
\cite{IK85} \cite{BDP92} at $m_q = .1$ predicts,
in fact, a strong first order transition at $\mu = .65$ ( The
small difference in $\mu_c$, can probably be accounted for 
by 1/d corrections ).
The analytic expressions of the strong coupling expansion show a 
feature not seen in the monomer--dimer simulations: 
a mixed phase for $\mu_o < \mu < \mu_s$
where ordinary confined hadronic matter coexists with 
the saturated lattice phase.

The quenched results \cite{LKS96} were characterized by a 
``forbidden region'' ranging from $\mu_o = m_\pi/2 = .32$ to
$\mu_s \simeq m_B/ 3 \simeq 1.0$.
$\mu_o$ and $\mu_s$ are close to the extrema of the
mixed phase predicted by the strong coupling expansions
mentioned above.  There is no remnant of the critical 
point for chiral symmetry restoration 
$\mu_c \simeq .65$ predicted by the same expansion.

Our motivation in undertaking the $6^4$ calculations, was, of course, 
to see results completely different from the 
quenched calculations
and very similar to the monomer--dimer results.

\begin{figure}
\epsffile{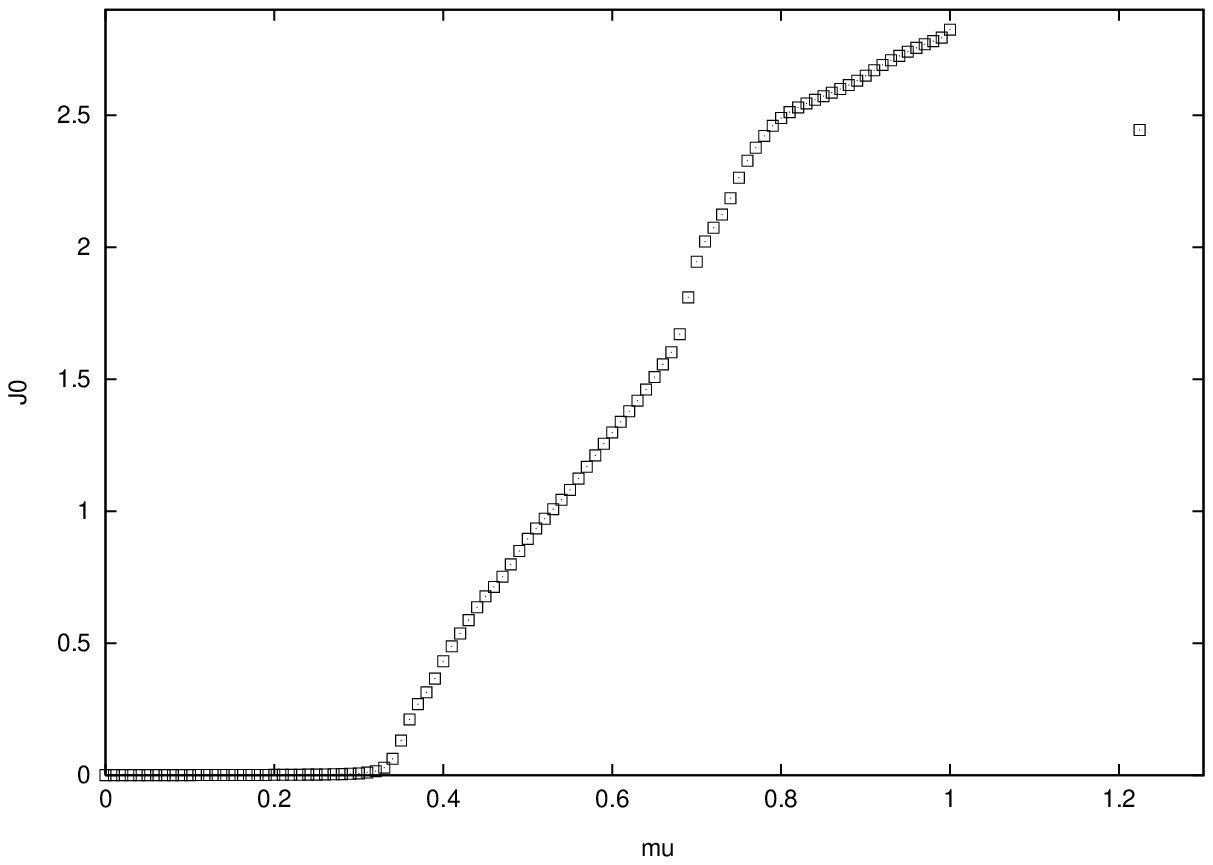}
    \caption[xxx]{Quark number density from the Glasgow algorithm.  
$m_q = .1$ on a $6^4$ lattice. The onset $\mu_o$,
and the saturation point $\mu_s$, are the same as the
ones observed in the quenched approximation. The critical
point for chiral symmetry restoration measured in a 
monomer--dimer calculation is $\mu_c = .69(1)$, coincident with the
little gap observed in our results. The same monomer--dimer results
would, however, predict a very sharp transition with a critical
density close to zero, in agreement with the results of the strong coupling
expansion.}
    \label{fig:numden}
\end{figure}

These expectations were only partially borne out. 
We indeed found a signature at $\mu_c$,
but {\it also} the persistence of $\mu_o$ and
$\mu_s$.
These results are shown in Fig. \ref{fig:numden},
for the number
density obtained with the Glasgow method on a $6^4$ lattice
at $m_q = .1$.
The Glasgow results
are distinguished from the quenched ones
by a small jump at $\mu \simeq .7 \simeq \mu_c$, 
suggesting restoration of chiral symmetry. Other than that, the Glasgow
and quenched results are very similar.

We then moved to a larger lattice 
to study the sensitivity to
size and temperature. Would the small hint
of a discontinuity at $\mu \simeq .7$ become more pronounced? Would the dynamical results
differ more substantially from the quenched ones?

The answer was in the negative. Neverthless, we did learn
something from these runs.

In Fig. \ref {fig:enlarged_1}, 
we show a detailed comparison of the
results on the two lattices, $6^4$ and $8^4$,
$m_q = .1$ (note the different scales
on the right and the left side). 
By blowing up the picture of the number density, we 
see that the density itself deviates from
zero at $\mu \simeq 0.$. This effect is very small (note
the  scale) and it is sensitive to temperature, 
the number density being 
suppressed, as expected, on the colder lattice. 

The most interesting point is that temperature effects are
greatly lessened for $\mu > \mu_o$. One might have well thought 
that the increase at $\mu_o$ reflects a thermal excitation of
baryons.  This does not seem to be  the case: 
only for $\mu < \mu_o$ do we observe the expected, physical pattern
of finite temperature effects. This disappears for $\mu > \mu_o$.
This result supports the belief that the rise at $\mu_o$ 
is unphysical, as in the quenched approximation. 
Of course we cannot rule out the possibility that the situation
changes on larger lattices, and we refer to \cite{KLS95} 
\cite{LKS96} for discussions on this point.

Temperature effects become 
apparent again at $\mu_c$,  
suggesting that $\mu_c$ is a threshold of a new phase.

\subsection{The determination of the critical point}

We can substanstiate this intepretation of $\mu_c$ 
by examining the zero's of the grand partition function in the
complex plane $e^{\mu}$.

The numerical strategy suggested by Section II, 
eq. \ref{for:jjj}
is straightforward:
observe the distribution of the modulos of the zeros, or,
equivalently,
search for a strip of high densities in the $e^\mu$ plane.
This criterion is numerically more convenient than the conventional Lee--Yang
analysis, which only uses the zero whose imaginary part is closest
to the real axes. 
It is also very natural : it says that the
number density counts the density of 
states in the fermionic sector.

\begin{figure}
\epsffile{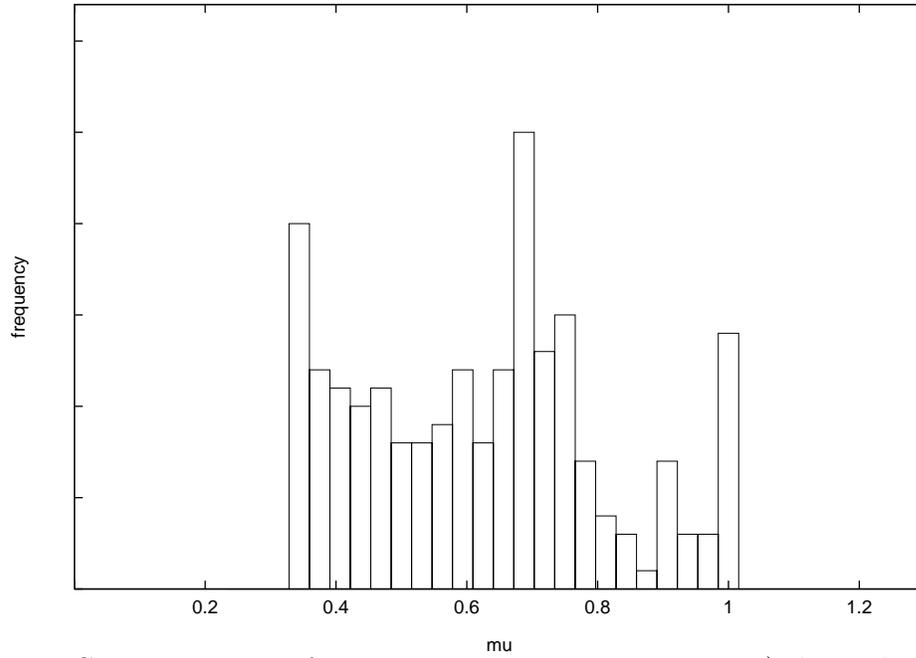}
\caption{Histogram of zeros accompanying
Fig. \ref{fig:numden}. Note a) the peak at $\mu_c = .687(15)$ matching
the small jump, to be contrasted
with the monomer--dimer results $\mu_c = .69(1)$. 
b) the correspondence of the extrema
of the histogram with the onset $\mu_o$ 
of the current and its saturation $\mu_s$. } 
\label{fig:fvq_n}
\end{figure}

This strategy is demonstrated in Fig. \ref{fig:fvq_n} where
we show the distribution of zeros 
(in practice, of the logarithms of their moduli)
accompanying Fig. \ref{fig:numden}.
The signal at $\mu_c = .687(15)$ is very clear, and in excellent
agreement with the monomer--dimer results $\mu_c = .69(1)$.

\begin{figure}
    \epsffile{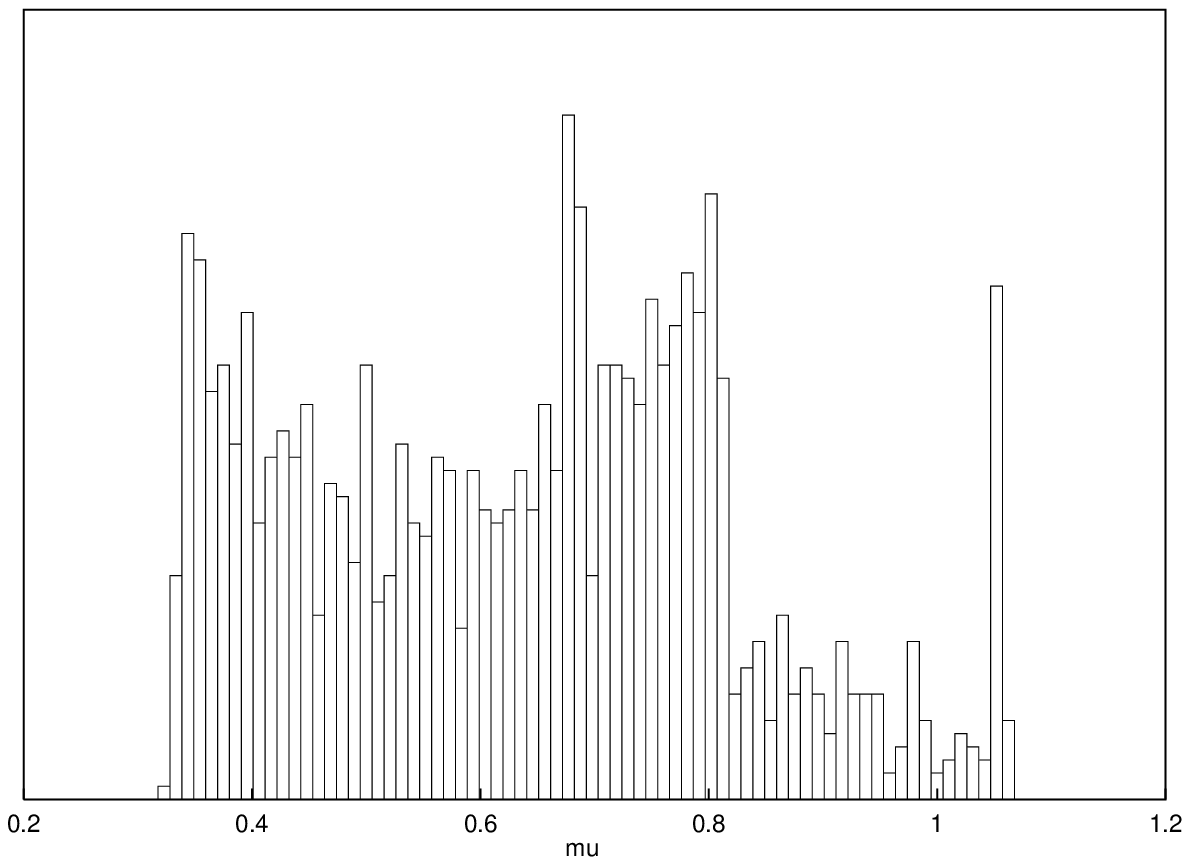}
       \epsffile{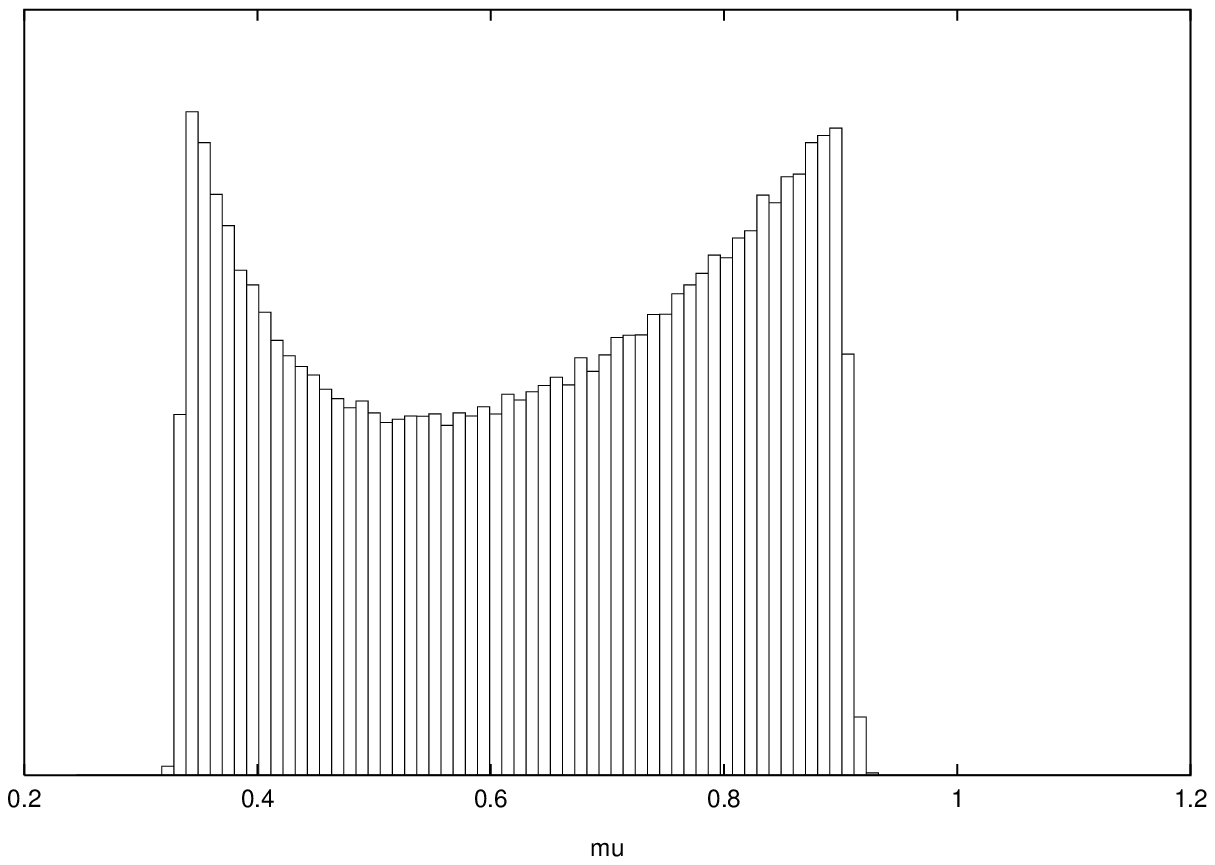}
\caption{ Histogram of the zeros
of the full model (top), 
and Histogram of the zeros of
the determinant (bottom), hence
of the quenched approximation. $m_q = .1$,
on a $8^4$ lattice.}
\label{fig:fvq_d}
\end{figure}

Figs. \ref{fig:fvq_d}
contrasts zeros of the determinant and zeros 
of the GCPF on a $8^4$ lattice. 
The signal at $\mu_c$ in the full model is quite clear.

As discussed
above, the upper and lower parts of the figure can also be seen as
full and quenched results. 
As a by--product of our investigation,
we see clearly why  the search of further 
critical points in the quenched calculations 
was futile \cite{Go88} \cite{LKS96}: the eigenvalue 
distribution is almost ``flat''. 

Unfortunately, the upper and lower parts of the figure also show
strong signals at $\mu_o$ and $\mu_s$:
the behaviour at the two side peaks does not
change as we pass from the quenched to the full model.
This gives us another view of the puzzling persistence
of the onset noticed in the previous subsection. The Glasgow simulation
method has failed to reproduce the published monomer--dimer results.

\begin{figure}
\epsffile{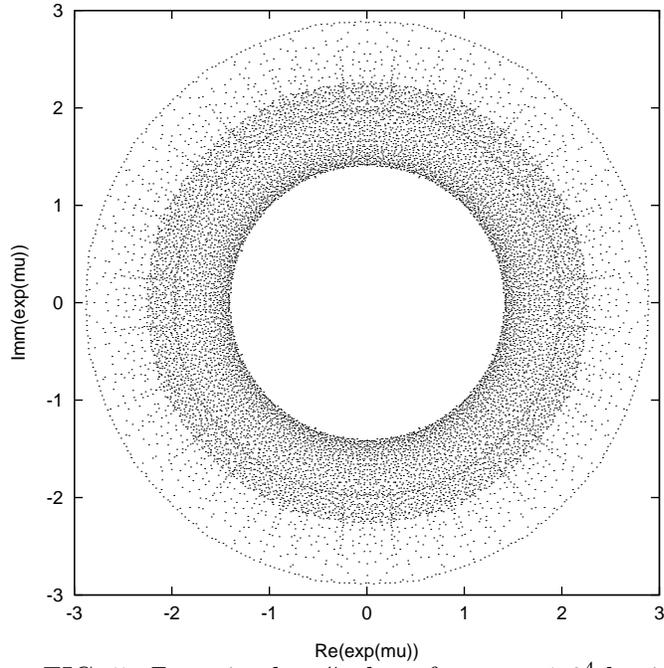}
    \caption[xxx]{%
Zeros in the $e^\mu$ plane 
for $m = .1$  $8^4$ lattice. The critical line is
the thin line inside the denser region $e^{\mu} = e^{\mu_c}$.}
\label{fig:zero}
\end{figure}

We can also look at the zeros themselves,
which we display in 
Fig. \ref{fig:zero}. As anticipated in the discussion, we observe a 
dense line, which follows
the prediction $|e^\mu| = e^{\mu_c}$. We also see a ring of zeros
at half the pion mass, and note that
the zeros fill up the entire region
$\mu_o$--$\mu_s$.

In summary, we have seen how the small discontinuity observed
in the current manifests itself in the histogram of zeros:
the density of zeros is the
{\it derivative} of the number density, so a small ``discontinuity''
in $J_0$ corresponds to a distinct signal in the histogram of zeros.

Will more statistics eventually cancel the 
onsets at $\mu_o$ and $\mu_s$? 
Even if we haven't observed any dramatic effect by increasing the
number of configurations, insufficient statistics
remains a possibility,
especially since $Z_3$ invariance has not been completely
achieved yet, and since it is possible that the precision required to
achieve the cancellation of the unwanted onset is prohibitively 
high. It is also possible that the polynomial representation for the
GCPF is ill-conditioned\cite{W59}.

\subsection{Heavy and light masses}

All the results we have discussed above were for
$m_q = .1$. It is instructive to explore
both light and heavy masses. Light masses are important
for the chiral limit. There we expect the critical
point $\mu_c$ to remain constant and different from zero.
Heavy masses change the critical point and allow a
more detailed comparison
of our results with the monomer--dimer/strong coupling approaches.

\begin{figure}
    \epsffile{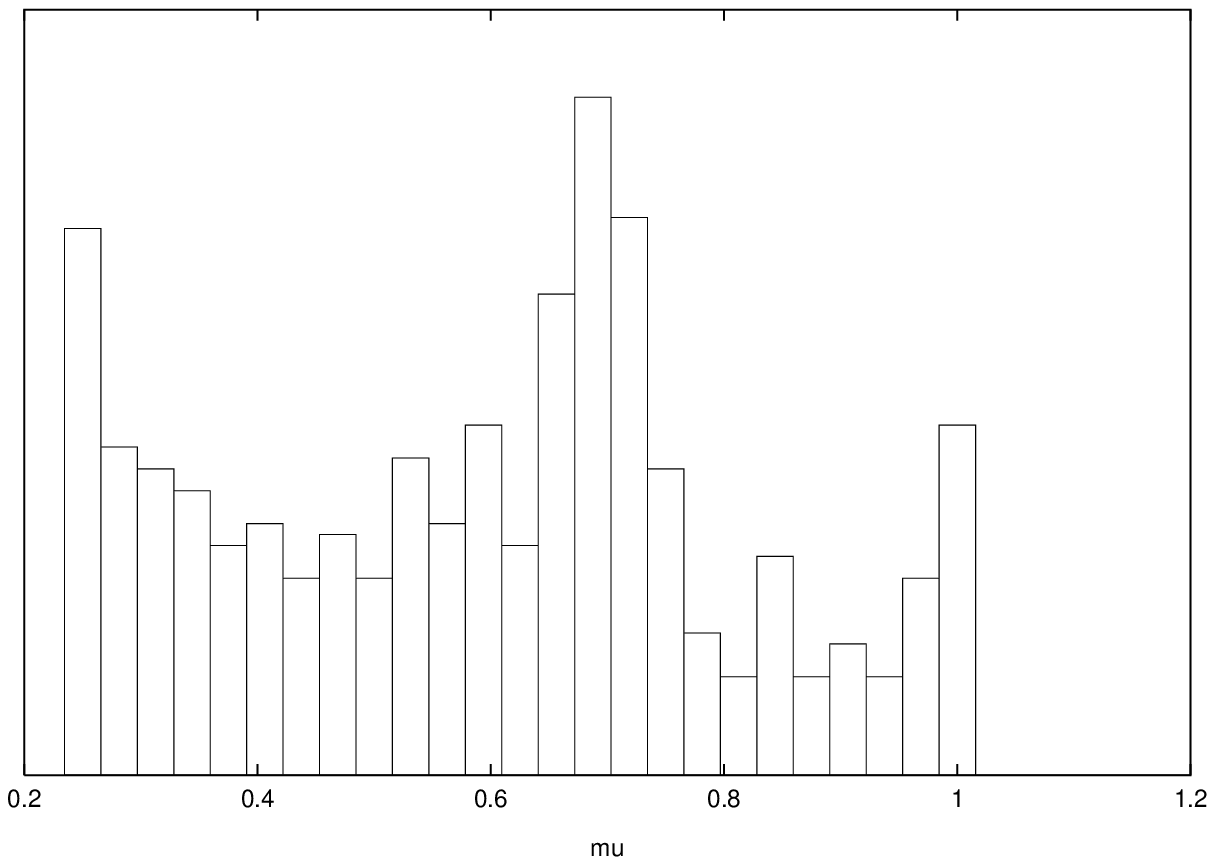}
       \epsffile{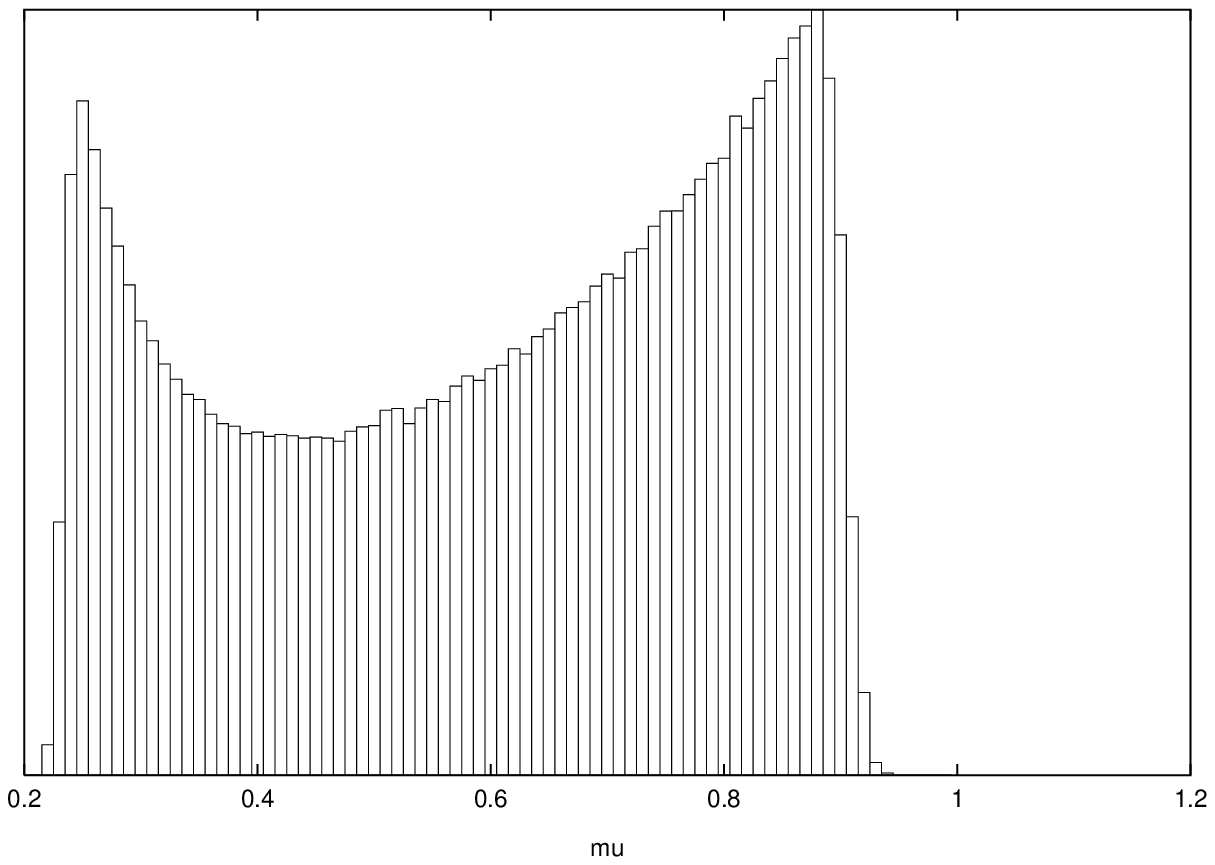}
\caption{ As in Fig. \ref {fig:fvq_d}, but $m_q = .05$, on a $6^4$ lattice.}
\label{fig:fvq_c}
\end{figure}

\begin{figure}
       \epsffile{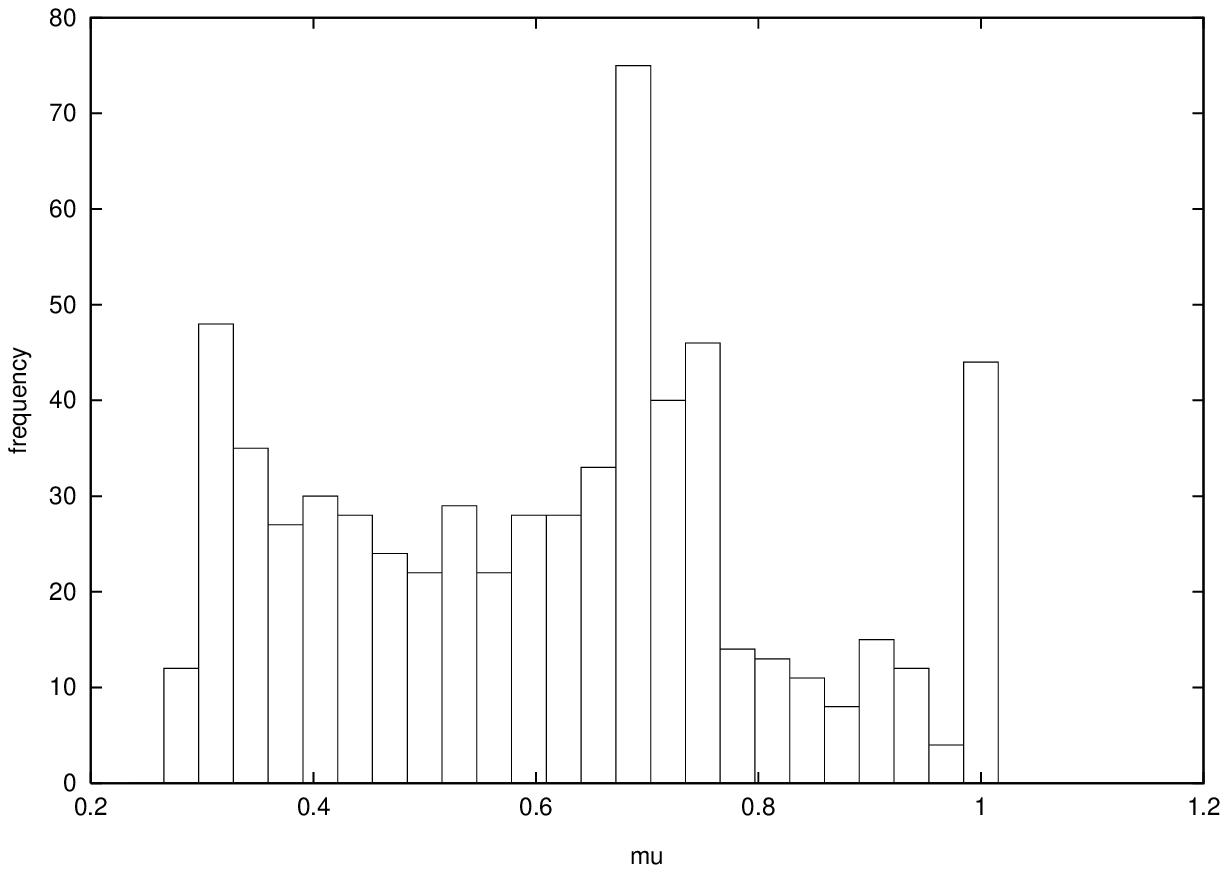}
       \epsffile{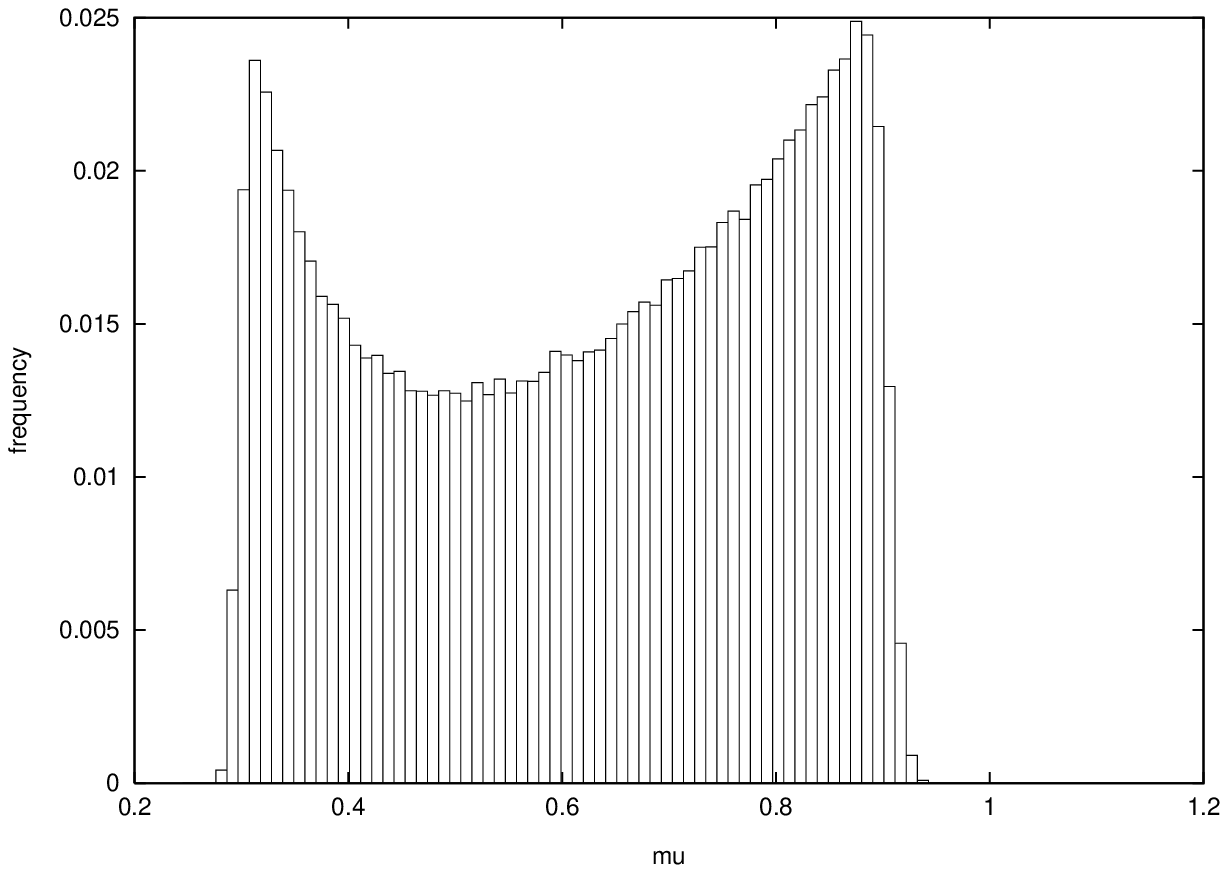}
\caption{ As in Fig. \ref {fig:fvq_d}, but $m_q = .08$, on a $6^4$ lattice.} 
\label{fig:fvq_b}
\end{figure}

Figs. \ref{fig:fvq_c}  , \ref{fig:fvq_b} 
show the sensitivity to the quark mass close to the chiral
limit : note the stability of the central peak, 
and the shift of the lower peak, which follows the onset 
of the current (see again also Fig. \ref{fig:fvq_d}).

\begin{figure}
    {\epsffile{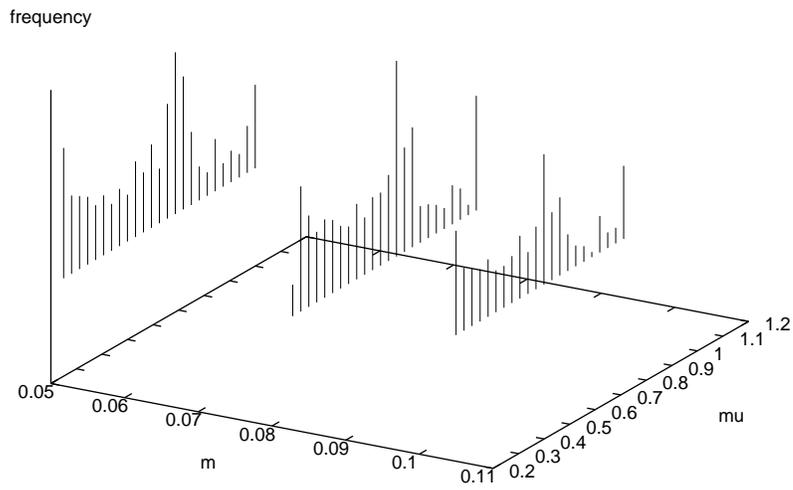} 
     \epsffile {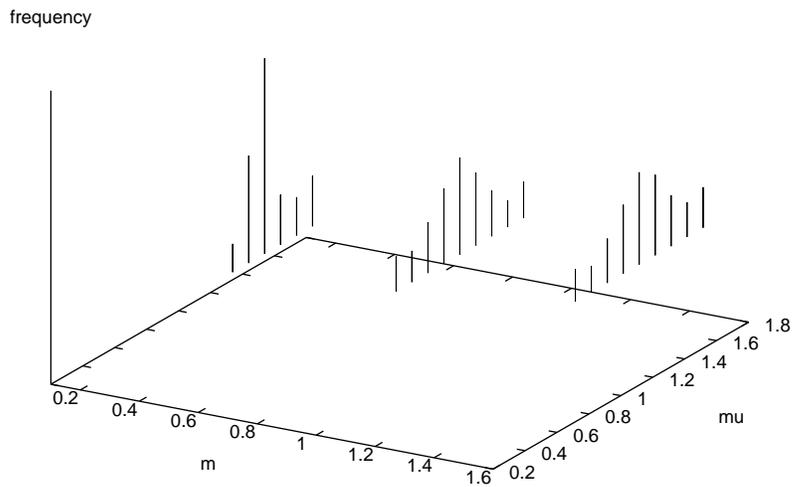}}
    \caption[xxx]{Overview of the histograms 
for various quark masses on a $6^4$ lattice.}
\label{fig:onsets_a}
\end{figure}

Finally,  we have also simulated larger masses,
and the results are displayed in Fig. \ref{fig:onsets_a}.
We can appreciate  
the shifting in the central peak, and also
its broadening -- probably due to the fact that at large
quark mass the transition is washed out. Interestingly, the
current onset at larger quark mass is, apparently, smaller
than half the pion mass -- a surprising result since
at $m_q = 1.5$  in the quenched model the critical region
shrinks to zero \cite{BBD86} -- certainly this adds to the complication
and the confusion associated with $\mu_o$.

However, even if the interpretation of the critical region at this
stage is largely subjective, the estimate of the critical point
$\mu_c$ seems reasonably sound.

\begin{figure}
  \epsfig{file=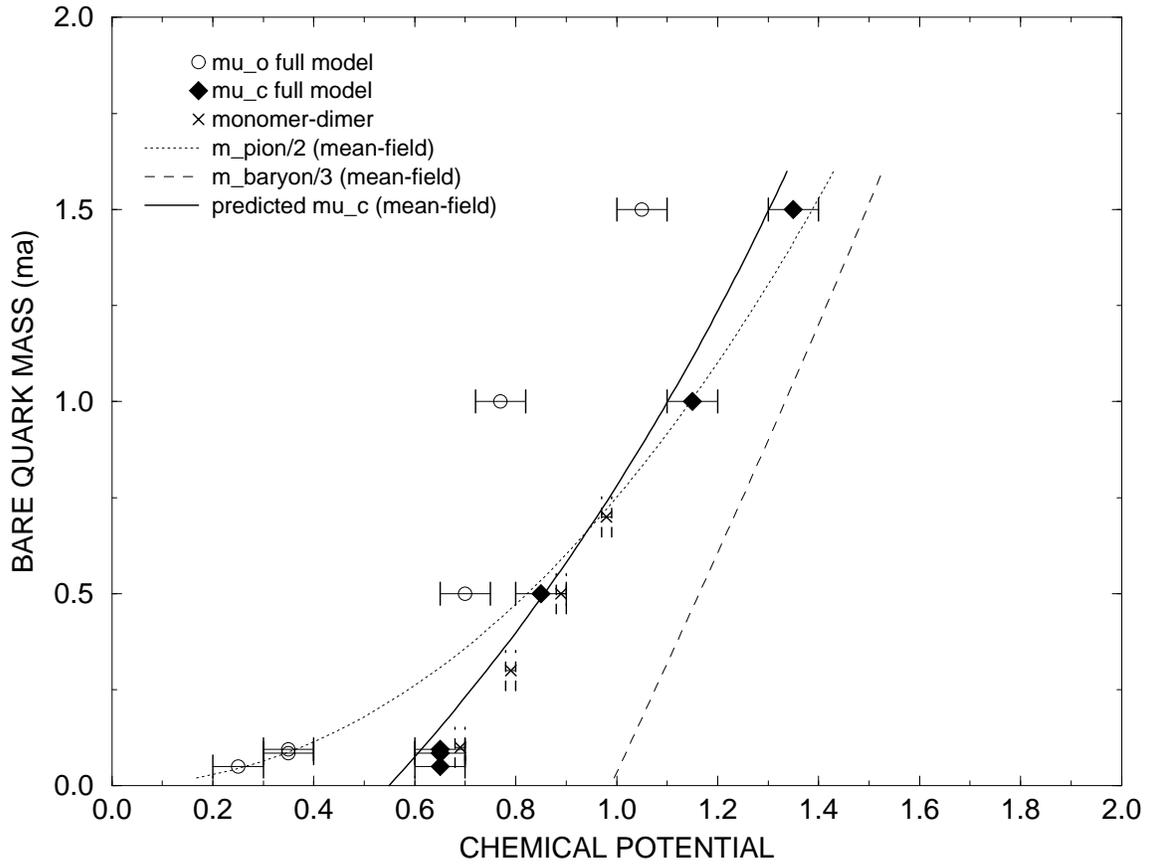, angle=-90}
\caption{Summary of the
results for the critical point $\mu_c$, and
current onset $\mu_o$. $\mu_c$ follows the prediction of the mean
field analysis of ref.[18] (solid line). 
The onset is close to half the pion mass at small mass, 
and below half the pion mass for $m_q > .5$ .}
\label{fig:susan_summ}
\end{figure}

We then conclude our results Section
with the summary of Fig. \ref{fig:susan_summ}.

\subsection{The analysis of
the probability distributions}

In this Section we re-examine the critical region by studying
the probability distribution for the particle number. 

Write
\begin{equation}
Z =  \sum_{n =-3 V}^{3 V} W_n 
\end{equation}
and normalize such that $Z = 1$.
$W_n$ is then the probability that a system in a grand canonical
ensemble has $n$ particles.

Using the numerical results for the GPF above (see eq.(\ref{for:basic}), 
the shapes of the probability distributions $W_n = Q_n e^{\mu n_t n}$
for different chemical potentials
can be drawn as a function of $n$, and the critical region can
be studied using standard statistical mechanics analysis.

For a transition in a classical ensemble 
in the infinite volume limit,
the distribution of $W_n$ should have a single peak in a pure phase,
and a flat distribution at the critical point, where
all values of the particle number
between the two extrema should be equally likely.

The current $<J_0> $ can be written as
\begin{equation}
<J_0>  = \frac{1}{V} \sum_{n = -3V}^{3V}n W_n
\end{equation}

\begin{figure}
{{\epsfig{file=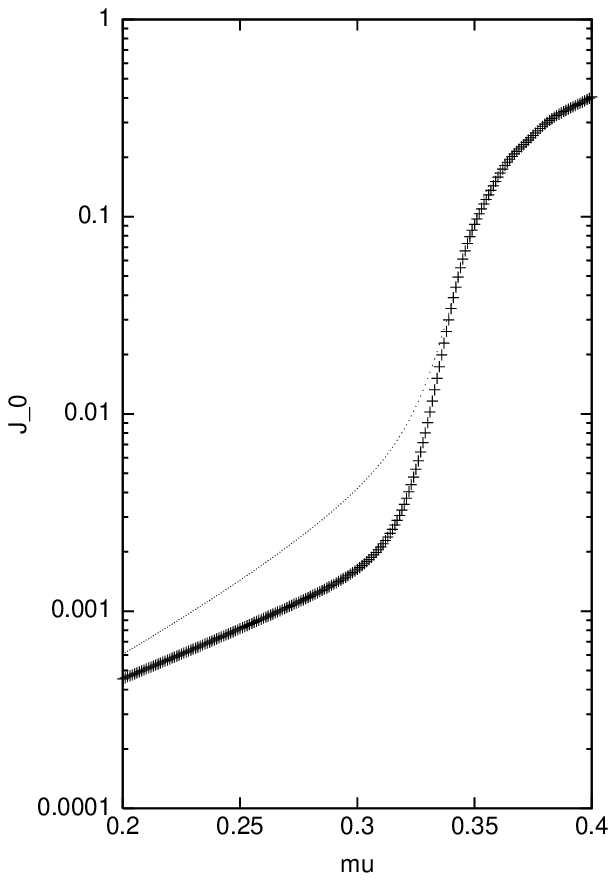, width = 7cm}} 
{\epsfig{file=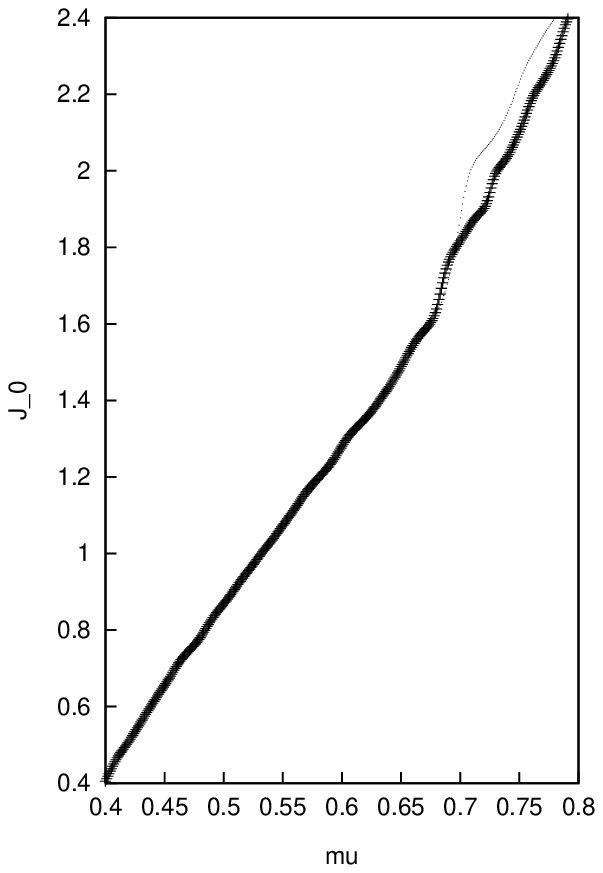, width = 7cm}}}
\caption[xxx]{Finite size effects at $m_q = .1$.
We show details of the critical regions around $\mu_o$ 
and $\mu_c$  for $m_q = .1$ for two different lattices.
The thick lines are for the $8^4$ lattice, the thin lines
for the $6^4$.}
\label{fig:enlarged_1}
\end{figure}

\begin{figure}
\epsffile{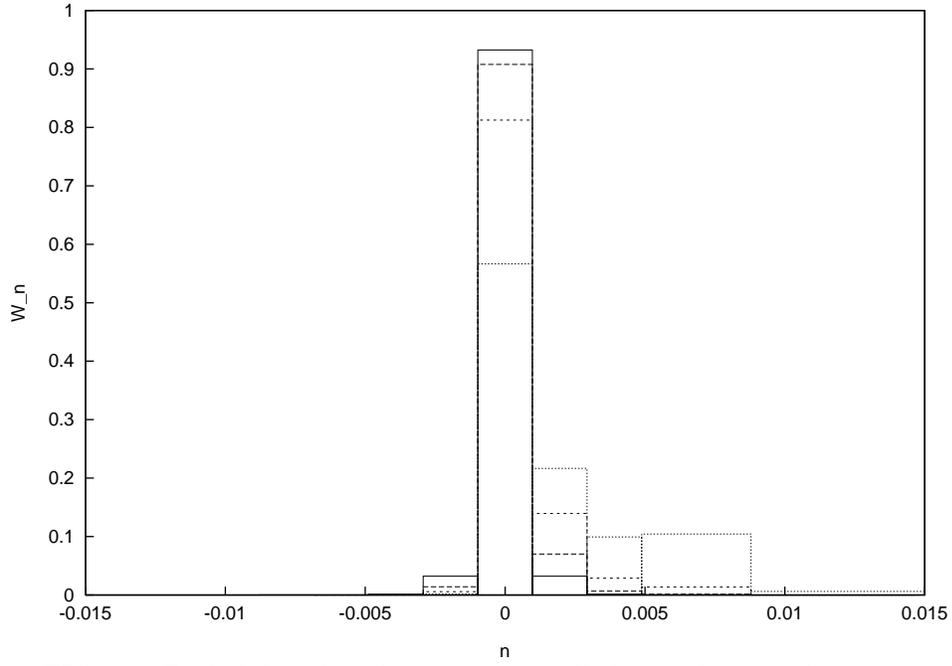}
    \caption[xxx]{Probability distributions for small chemical
potentil at $m_q = .1$ on the $8^4$ lattice. The solid line is
$\mu = 0$, the dashed lines , from top to bottom at
$n=0$, are for $\mu =  .1 , .2 , .3$.}
    \label{fig:wn_small_mu}
\end{figure}

We draw in Fig. \ref{fig:wn_small_mu} the probability
distribution for small chemical potential. 
At $\mu = 0.$ (solid line) the distribution
is symmetric around the origin : look
at the two satellite peaks of equal height. $<J_0>$ equals zero
as it should. 
At $\mu = .1$ (dashed line) the distribution becomes asymmetric,
reflecting the enhancement (suppression) of the 
forward (backward) propagation : the peak on the left
decreases, the one on the right increases.
Positive and negative states are
still both contributing to the probability distribution. The net
$<J_0>$ moves immediately off zero, but it is very, very small
(look again at the picture \ref {fig:enlarged_1} ).
The distribution broadens on smaller lattices, which accounts
for the pattern of finite size effects seen in the same picture.

\begin{figure}
\epsffile{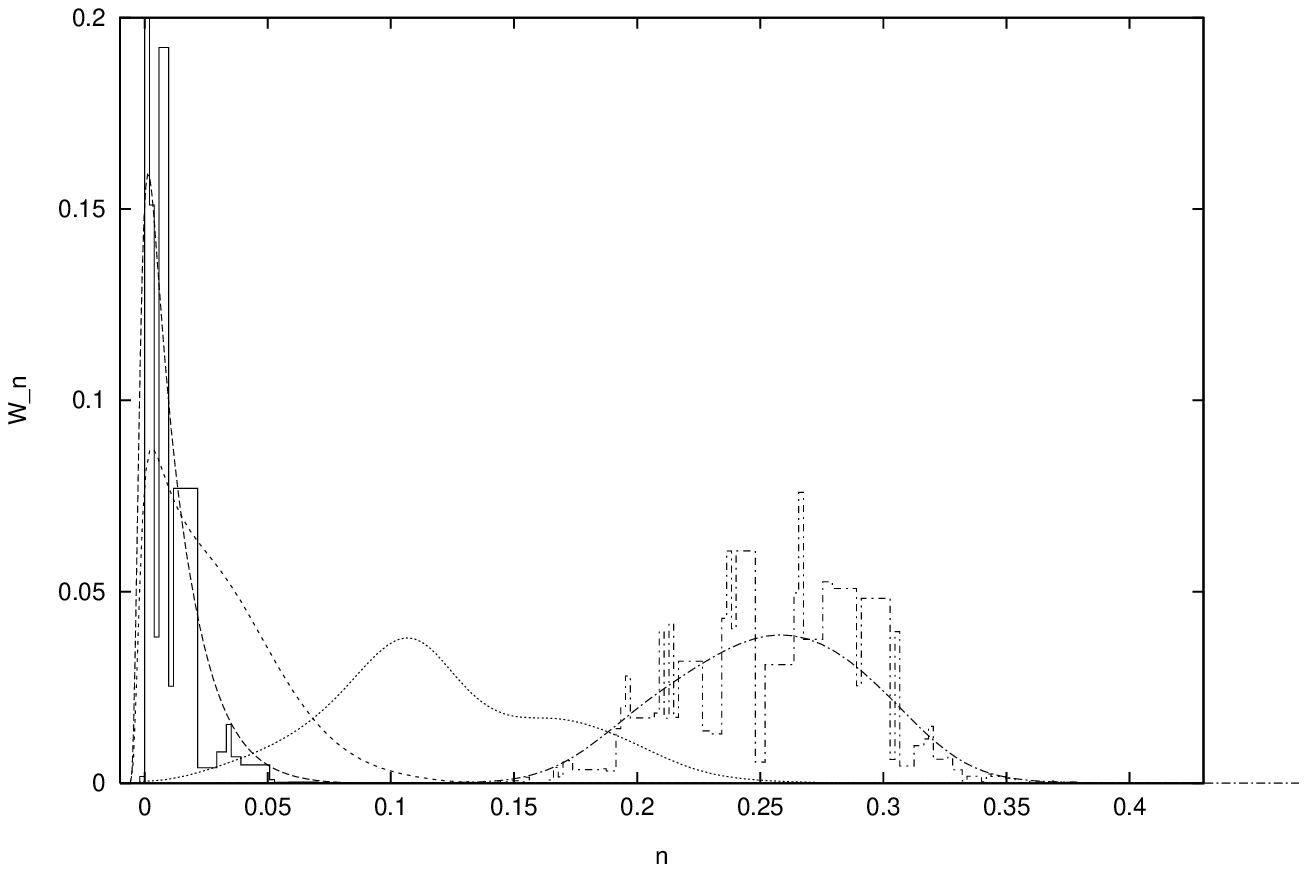}
\epsffile{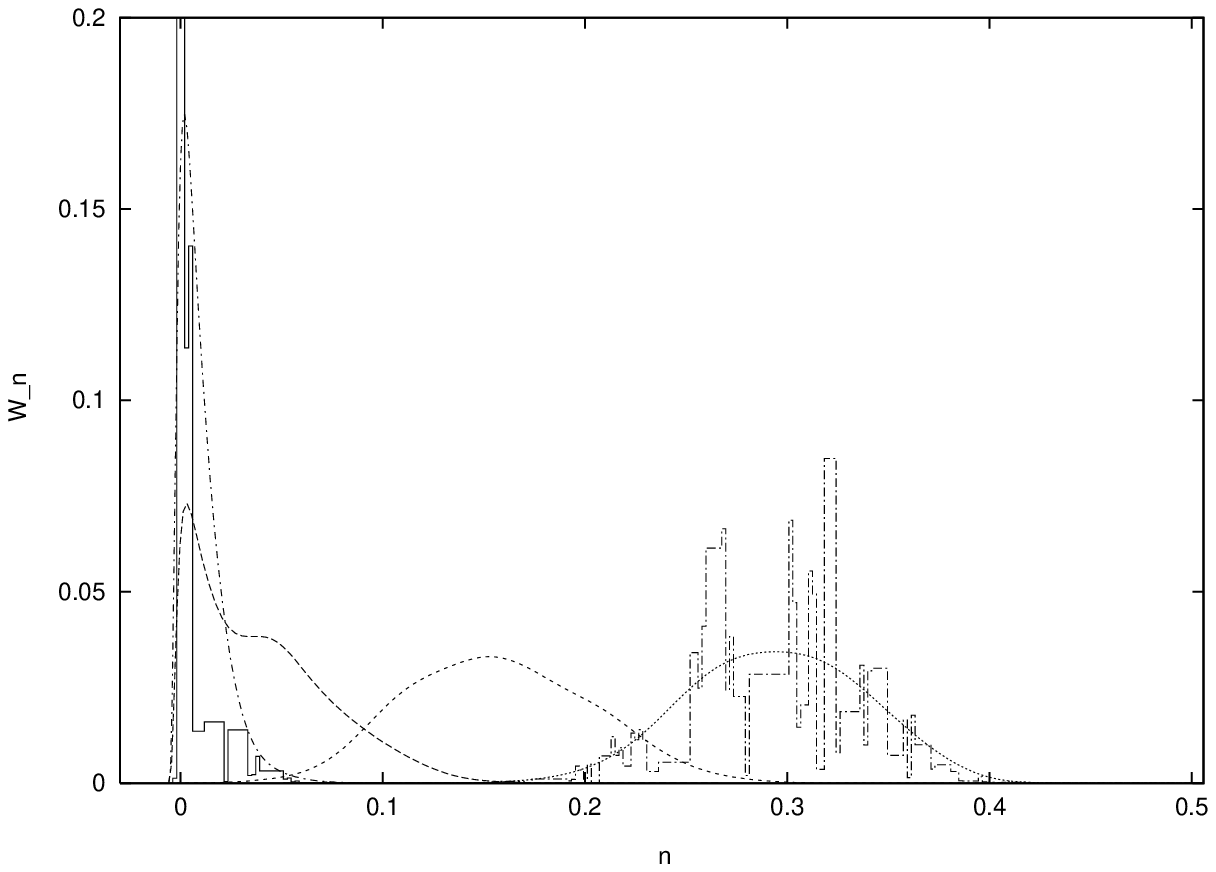}
    \caption[xxx]{Probability distributions around 
the onset $\mu_o$ for $m_q = .08$ (top) and
$m_q = .1$ (bottom) on a $8^4$ lattice. 
The leftmost histogram (solid) at $m_q = .08$
is for $\mu = .28$ , the rightmost is for $\mu = .34$.
Bezier interpolations (from Gnuplot) are shown for
$\mu = .28, .30, .32, .34$ . At $m_q = .1$
$\mu = .32, .34, .36, .38$ from left to right. For both
masses at $\mu_o$ the probability distribution moves on the positive
$n$ axes.}
\label{fig:wn_onset}
\end{figure}

At $\mu = \mu_o$ the scenario changes completely: a secondary maximum
develops at positive $n$, and the distribution moves to the positive 
$n$ region. We show this behaviour for both $m_q = .1$ and
$m_q = .08$ in Fig. \ref {fig:wn_onset}. For $\mu > \mu_o$ the
negative states do not contribute.
This behavior is correlated with the sharp increase of $J_0$
plotted before and should be related to changes in the
theory's spectrum, perhaps reflecting pathologies of
the quenched case such as the `` funny pions''  \cite{LKS96} 
or Stephanov's condensates \cite{MIS96}.
The distribution is now roughly symmetric, and its broadening on 
smaller lattices does not affect its average value
$<J_0>$. This behavior is compatible with the absence of strong 
volume effects, Fig.\ref {fig:enlarged_1}. 

\begin{figure}
\epsffile{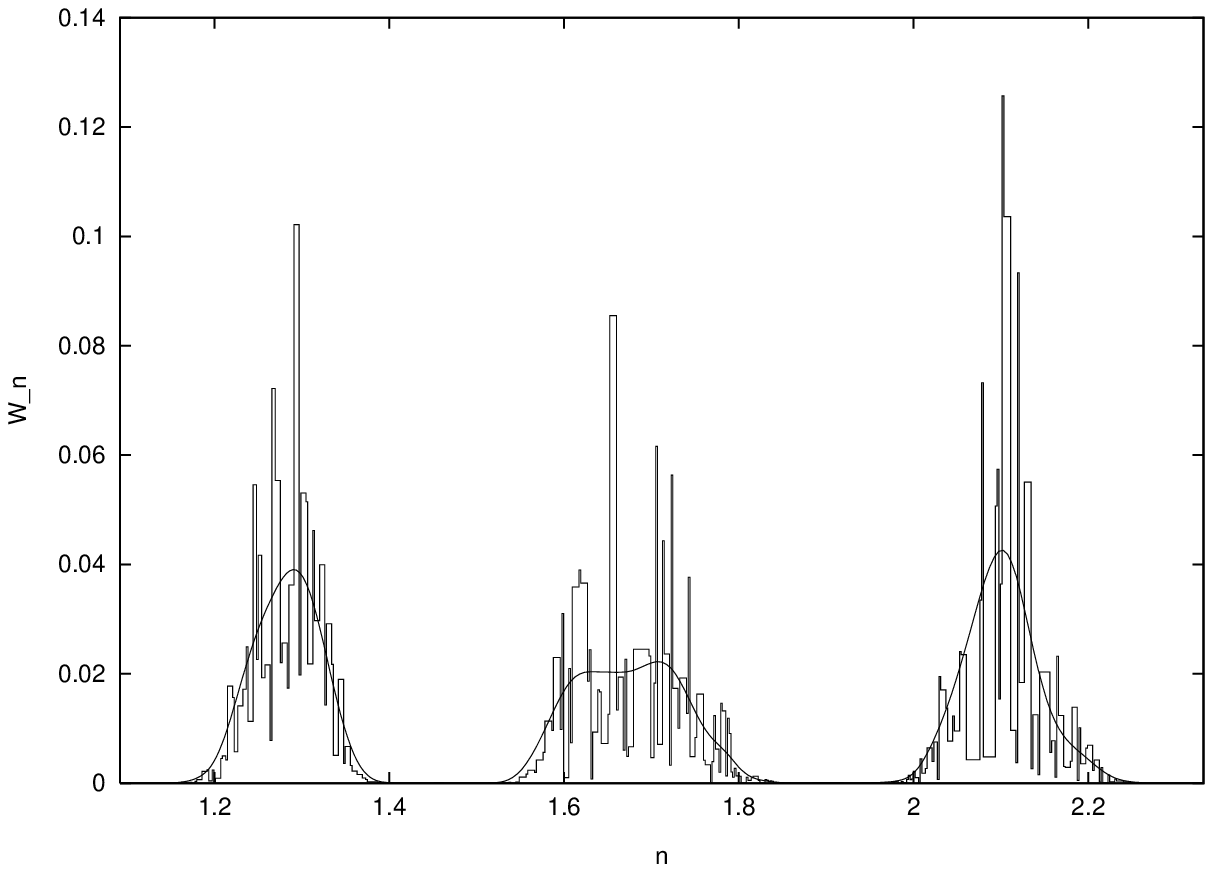}
\epsffile{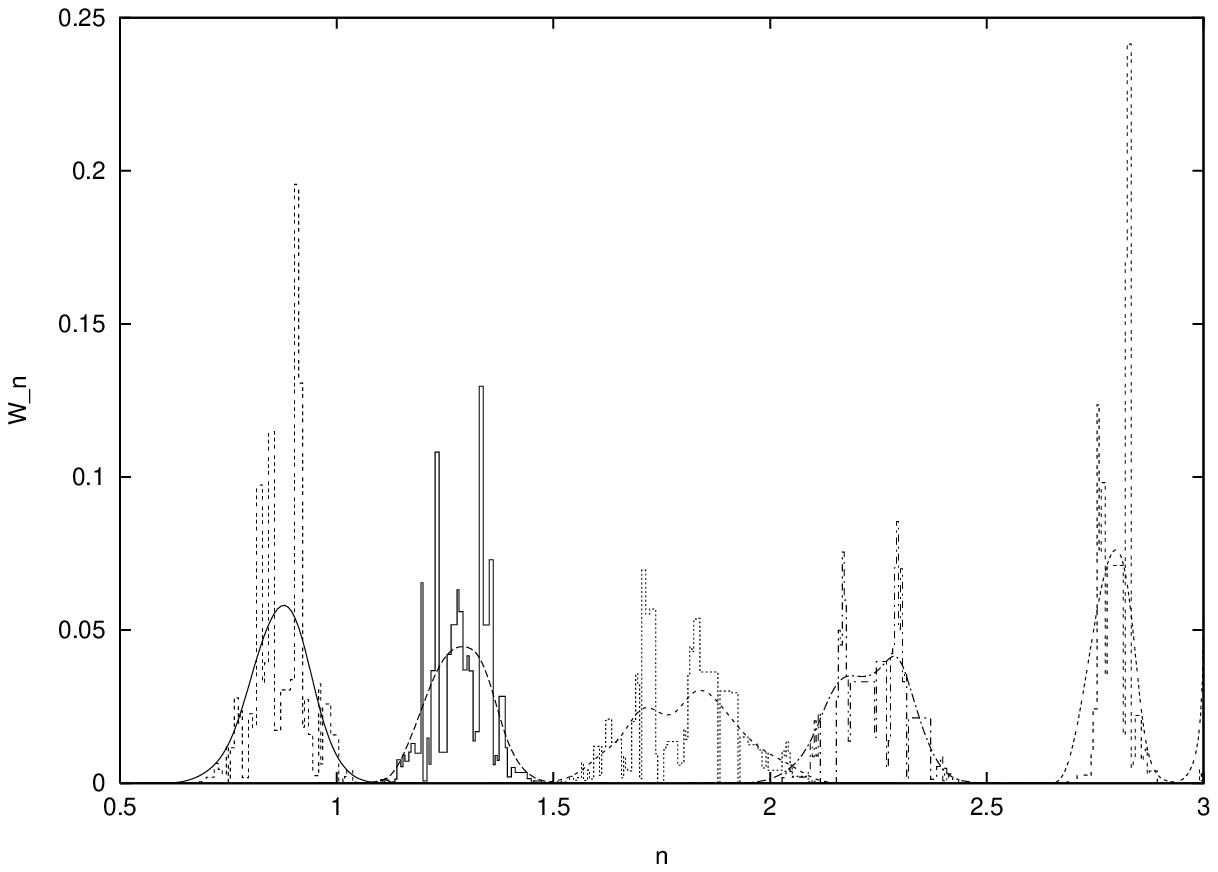}
    \caption[xxx]{Probability distributions in the critical
region at $m_q = .1$, on the $8^4$ (top) and the $6^4$ (bottom).
$\mu$ is (.6, .683, .75) , from left to right (top) , and
(.5. ,.6, .695, .75 , 1.) bottom. }
    \label{fig:wn_mu_c}
\end{figure}

\begin{figure}
\epsffile{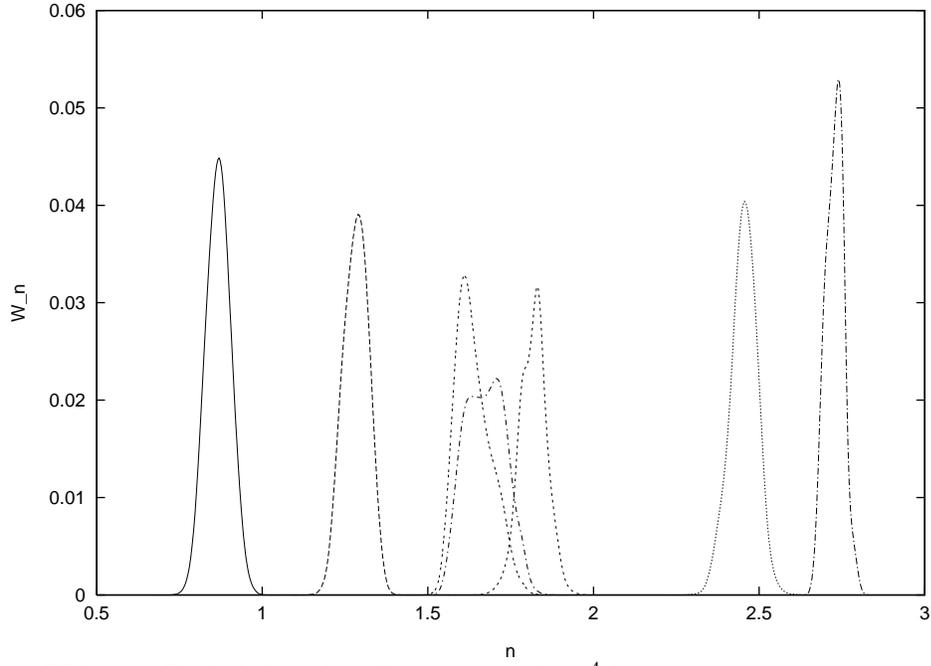}
    \caption[xxx]{Probability distributions 
on the $8^4$ lattice, $m_q = .1$, for 
$\mu$ = (.5 , .6 , .68, .683, .7 , .8 , .9). 
Only the Bezier interpolations are shown. The complete
results for several $\mu$ values can be seen in 
Fig. \ref {fig:wn_mu_c}. }
\label{fig:wn_mu_c_1}
\end{figure}

Next, the critical region : we see
the expected broadening
of the probability distribution at $\mu_c$ (Fig. 
\ref {fig:wn_mu_c} , \ref {fig:wn_mu_c_1}). 
Finite size effects
become important again for $\mu > \mu_c$. 
Note, in particular, in Fig. \ref {fig:wn_mu_c_1}, the sensitivity
to $\mu$ on a very fine scale : the three central plots are
for $\mu = .68, .683, .7$. 

\begin{figure}
\epsffile{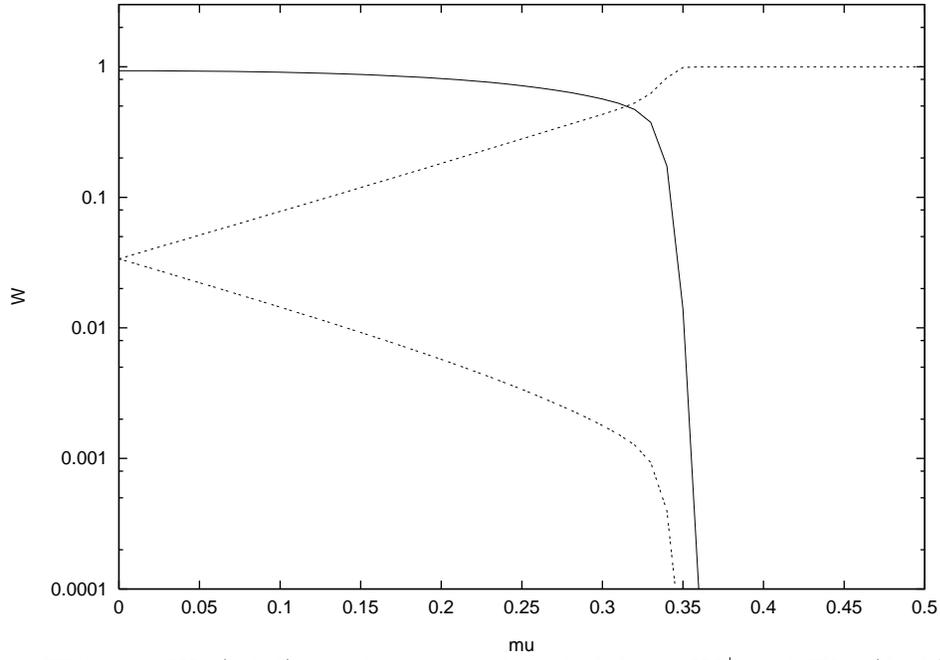}
    \caption[xxx]{$W_0$ (solid) , and integrated probabilities $W^+$ and
$W^-$ (dash) at $m_q = .1$, on the $8^4$ lattice}
    \label{fig:wn_summary}
\end{figure}

In Fig. \ref {fig:wn_summary} we summarize these observations
by plotting $W_0$, the probability that the system has 
zero particle number,  
and the integrated probabilities $W^+ = \sum W_n , n > 0$;
$W^- = \sum W_n , n < 0$.
The logarithmic scale of the plot makes it easy 
to see that backward and forward propagations are enhanced
and suppressed by the same factor at small $\mu$. Correspondingly, the
contribution of $n=0$ must decrease.
At $\mu = \mu_o$ $n=0$ equals the overall contribution from $ > 0$.
For $\mu > \mu_o$ only positive $n$ contribute to $Z$.

These results suggest that
$\mu_o$ is the threshold for a phase with only positive
propagation. Perhaps this observation is a clue to the nature
of the phase $\mu > \mu_o$ .  Recall that mean field analysis predicts 
the threshold of the mixed phase (broken phase/
saturated phase) at $\mu \simeq \mu_o$. 
Future work should address possible
relations between these observations.

We believe that $\mu_c$ indicates a physical critical point.
All approaches (except the pathological quenched case) predict a transition
or, at least, a clear change of behavior of observables here.
From the point of view of this section it is relatively easy to
understand the robustness of this result :
the probabilities plotted here underlie all the observables 
discussed earlier and the ``flattness'' of the
distributions, which locates the critical point, 
is a qualitative feature which should appear in
all the numerical procedures.

\section{Outlook}

We have reached a partial understanding
of the algorithms and we can point out some successes and failures.

On the positive side, the method gives clear signs of the critical point 
$\mu_c$ which should be the point of chiral symmetry restoration.
The method also gives a current onset $\mu_o$ far different from $\mu_c$. 
This is most likely unphysical: it is not seen in the
monomer--dimer results, it is the same as the pathological quenched onset, and
it is the threshold of a phase characterized by forward propagation. 
Unfortunately, 
since $J_0$, and other observables such as the energy density, deviate from
from zero at the unphysical $\mu_o$ point, their values near $\mu_c$ cannot
be trusted. So, although the present algorithm gives 
$\mu_c$ accurately, it does
not make any other phenomenologically reliable predictions.

We expect that the early onset, $\mu_o$, should disappear in a correct calculation.
Physical arguments support this view as well as the
monomer--dimer and strong coupling expansions discussed here.
It might be that a high statistics run of the present algorithm will cancel $\mu_o$. In this case
the method would be impractical, but, at least, not
conceptually wrong. If this were true, we should develop
a strategy to monitor the convergence of the method
to the correct statistical ensemble, and to remove unphysical
contributions to observables due to partial
``cancellations'' of unwanted onsets.

A very unpleasant possibility, which we cannot exlude a priori, is that
the results we are observing are indeed the final results at finite
chemical potential with the Glasgow method. 
In this case, monomer--dimer simulations and strong coupling calculations
would differ from the Glasgow results.
This result would indicate intrinsic 
difficulties of the finite density lattice gauge theory
simulation strategies. Some of these have been discussed in the text. 

We might also suspect that the problems stem
from having generated configurations at zero chemical potential.
This explanation is suggested from the standard problems encountered
by reweighting procedures, and from the behaviour observed in 
the Gross Neveu model \cite{BMKKL97}. 
In this case the Glasgow method can be improved if a better
starting point were invented. This is a 
worthwhile direction to pursue.

At the present point, we have to accept that ensemble
averaging does not help to suppress the pathologies
of isolated configurations.
It might well be that a satisfactory  simulation of 
finite density QCD
requires an algorithm which produces physical results
on each configuration.
A promising development of this sort is 
$\chi$QCD \cite{DKK}, where an irrelevant four fermi 
term is added to the 
standard QCD action used here. 
$\chi$QCD has the advantage that chiral symmetry breaking
and the generation of a dynamical quark mass occurs 
configuration-by-configuration 
and the pion and sigma excitations are explicitly 
free of $\mu$ dependence. In fact,
$\chi$QCD simulations do not suffer from the 
severe $\mu_o$ pathologies seen here 
\cite{BMK96}, but additional work, both theoretical and practical, 
is needed to  see if $\chi$QCD really produces only physical results. 
Research in this topic is in progress.

\vskip .5 truecm
We wish to acknowledge valuable discussions with
N. Bili\'c, F. Karsch,  P. Kornilovitch, 
G. Parisi, H. Satz and M. Stephanov.
MPL would like to thank the Physics Department 
of the University of Bielefeld for its
hospitality,  and the High Energy Group of HLRZ/J\"ulich,
particularly K. Schilling, for support during the initial
stages of this project. 
The calculations were done at  HLRZ/KFA J\"ulich and 
at King's College, London.
This work was partially supported by Nato  grants nos. CRG950896 
and CRG960002 and by PPARC under grant GR/K55554.
EGK is supported by EPSRC grant GR/L03026.
JBK is partially supported by the National Science Foundation,
NSF-PHY92-00148.

\newpage
\newpage
\newpage
\newpage
\newpage
\newpage
\newpage
\newpage
\newpage
\newpage
\newpage
\newpage
\newpage

\end{document}